\documentclass[aps,prl,twocolumn,superscriptaddress]{revtex4-1}
\usepackage{graphicx}

\begin{document}

\title{Kinetic Signatures and Intermittent Turbulence in the Solar Wind Plasma}

\author{K.T. Osman}
\email{kto@udel.edu}
\affiliation{Centre for Fusion, Space and Astrophysics; University of Warwick, Coventry, CV4 7AL, United Kingdom}
\affiliation{Bartol Research Institute, Department of Physics and Astronomy, University of Delaware, Delaware 19716, USA}

\author{W.H. Matthaeus}
\affiliation{Bartol Research Institute, Department of Physics and Astronomy, University of Delaware, Delaware 19716, USA}

\author{B. Hnat}
\author{S.C. Chapman}
\affiliation{Centre for Fusion, Space and Astrophysics; University of Warwick, Coventry, CV4 7AL, United Kingdom}

\date{\today}

\begin{abstract}

A connection between kinetic processes and intermittent turbulence is observed in the solar wind plasma using measurements from the Wind spacecraft at 1 AU. In particular, kinetic effects such as temperature anisotropy and plasma heating are concentrated near coherent structures, such as current sheets, which are non-uniformly distributed in space. Furthermore, these coherent structures are preferentially found in plasma unstable to the mirror and firehose instabilities. The inhomogeneous heating in these regions, which is present in both the magnetic field parallel and perpendicular temperature components, results in protons at least 3--4 times hotter than under typical stable plasma conditions. These results offer a new understanding of kinetic processes in a turbulent regime, where linear Vlasov theory is not sufficient to explain the inhomogeneous plasma dynamics operating near non-Gaussian structures.

\end{abstract}


\maketitle

\textit{Introduction}.---A plasma turbulence cascade \citep{Coleman68,OsmanEA11b} might provide the energy needed to heat the lower solar corona \citep{CranmervanBallegooijen07}, accelerate fast and slow wind streams, and account for the non-adiabatic expansion of the solar wind \citep{Hollweg86}. However, turbulence at kinetic scales remains an unsolved problem and it is not known how fluctuation energy is ultimately converted into heat. There are a number of kinetic processes, such as linear wave damping \citep{Barnes79} and pressure-anisotropy instabilities \cite{Gary93}, that might play a role in this dissipation of the interplanetary cascade. These are often investigated using linear and quasi-linear approximations to the Vlasov-Maxwell equations. Other studies have suggested solar wind discontinuities are linked to turbulence intermittency, in the sense of coherent structures and non-Gaussian probability distribution functions \citep{GrecoEA09}, and are also sites of enhanced temperatures \citep{OsmanEA12}. Here we ask if kinetic effects are homogeneous in space, or concentrated in regions of the turbulent field associated with discontinuities. We find a link between magnetic coherent structures and kinetic signatures usually associated with linear instabilities.

\textit{Kinetic Physics}.---The kinetic plasma properties of solar wind proton populations have been studied in detail using \textit{in-situ} spacecraft measurements \citep{Marsch06}. Where samples have been subject to many collisional effects \citep{KasperEA08}, the proton velocity distribution functions (VDF) are typically isotropic and Maxwellian. However, less collisional solar wind plasma usually exhibits anisotropic VDFs with respect to the local magnetic field direction, such that $R \equiv T_{\perp}/T_{\parallel} \neq 1$. Solar wind temperature anisotropy $R$ is studied by investigating the dependence on proton parallel beta $\beta_{\parallel} = n_{p}k_{B}T_{\parallel}/(B^{2}/2\mu_{0})$, which is the ratio of parallel pressure to total magnetic pressure. These parameters cannot assume arbitrary values in the solar wind plasma, but are instead confined such that the range of observable $R$ is narrowed as $\beta_{\parallel}$ increases. This restriction could be linked to large deviations from $R = 1$ which can give rise to anisotropy-driven instabilities. These are not specific to the solar wind, and can operate in other low collisionality astrophysical plasmas. As the electromagnetic fluctuations associated with these instabilities grow, they are able to scatter particles and eventually drive the VDF back towards isotropy. From linear Vlasov theory, the solar wind plasma can become unstable to the mirror and cyclotron instabilities when $R > 1$, and to the firehouse instability when $R < 1$ and $\beta_{\parallel} \geq 1$. These calculations were compared with measured distributions of $R$ with respect to $\beta_{\parallel}$, and the data was found to be best constrained by the mirror instability for $R > 1$ and the oblique firehose instability for $R < 1$ \citep{HellingerEA06,KasperEA02}.

The solar wind plasma is found to migrate with increasing heliocentric distance towards higher $\beta_{\parallel}$ and to values of $R$ approaching unity \citep{MatteiniEA07}. At least part of this evolution is better understood by ordering the data according to collisional age $\tau = \nu_{pp}L/v_{sw}$, defined as the Coulomb proton-proton collision frequency multiplied by the transit time from the Sun to 1 AU, which is the number of thermalization timescales that elapsed on transit from the Sun to the spacecraft \citep{KasperEA08}. For low collisional age plasma, enhancements in power associated with magnetic fluctuations exist near instability thresholds \citep{BaleEA09} which also correspond to sites of elevated proton heating \citep{LiuEA06}. Indeed, plasma unstable to the mirror or firehose instabilities were found to be significantly hotter than stable plasma \citep{MarucaEA11}. These associations suggest underlying physical relationships between heating mechanisms in the solar wind and the kinetic physics that emerges from, or leads to, the growth of these instabilities. However, the details of these relationships are not completely understood. This Letter reports results linking instability thresholds and elevated temperatures to coherent structures dynamically generated by strong MHD turbulence. 

\begin{figure*}[t]
\includegraphics[width=17cm]{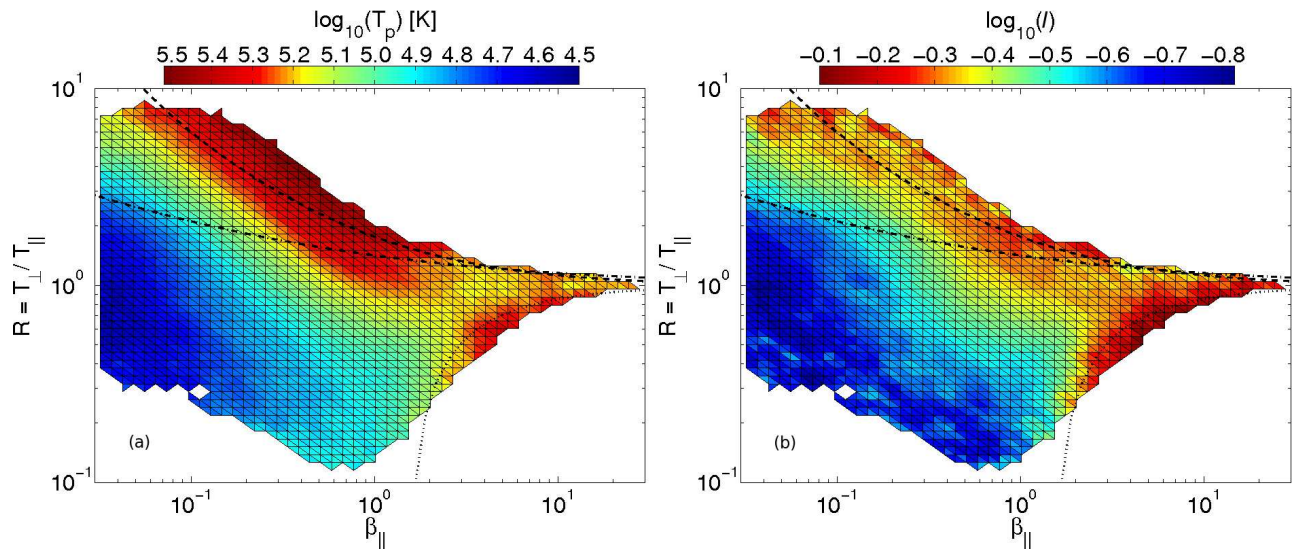}
\caption{Plot of median (a) scalar proton temperature $T_{p}$, and (b) PVI statistic $\mathit{I}$ over the $\left( \beta_{\parallel},T_{\perp}/T_{\parallel} \right)$-plane. The curves indicate theoretical growth rates for the mirror (dashed), cyclotron (dot-dashed), and oblique firehose (dotted) instabilities. There is a manifest association between these thresholds, hot plasma, and enhanced PVI.}
\label{Fig:PVIDist}
\end{figure*}

\textit{Analysis}.---We use around $2.5\times 10^{6}$ independent plasma and magnetic field measurements from the Wind spacecraft, recorded during the interval 1995 Jan. 1 to 2004 Nov. 20. The Faraday cup instrument of the Solar Wind Experiment (SWE) \citep{OgilvieEA95} measures proton density $n_{p}$, bulk velocity $\mathbf{v}_{sw}$, and proton temperatures. These are separated into parallel temperature $T_{\parallel}$ and perpendicular temperature $T_{\perp}$ by comparison with the mean magnetic field from the Magnetic Field Investigation (MFI) \citep{LeppingEA95}. Only solar wind data is used, and measurements either in the magnetosphere or contaminated by terrestrial foreshock are removed. We also require the uncertainties in the thermal speeds to be $ < 10\%$ and, to avoid Coulomb relaxation effects, the collisional age must be $\tau \leq 0.1$. These criteria follow \citep{MarucaEA11}, and only about 28\% of our original dataset satisfy all these conditions. 

Here we ask if kinetic effects, such as temperature anisotropy and heating, are strongly inhomogeneous and related to the intermittent character of the turbulent magnetic field. A way to find regions of high magnetic stress and coherent structures is to identify rapid changes in the magnetic field vector:
\begin{equation}
\Delta \mathbf{B}(t,\Delta t) = \mathbf{B}(t + \Delta t) - \mathbf{B}(t)
\end{equation}
where $\mathbf{B}(t)$ is the magnetic field time series and $\Delta t$ is the time lag. The fastest available cadence of the plasma data defines the lag $\Delta t = 92$\,s which, using Taylor's hypothesis, corresponds to a spatial separation in the plasma frame ($\Delta r = -v_{sw}\Delta t$) on the scale of inertial range fluctuations. We use the partial variance of increments (PVI) method to identify coherent (non-Gaussian) structures in the solar wind \citep{GrecoEA09}, where the highest PVI amplitudes are found at the smallest scales \citep{ServidioEA11}:
\begin{equation}
\mathit{I} = \frac{\left| \Delta\mathbf{B} \right|}{\sqrt{\langle \left| \Delta\mathbf{B} \right|^{2} \rangle}}
\end{equation}
where $\langle \cdots \rangle$ denotes an appropriate time average over many correlation times. Events are selected by imposing thresholds on the $\mathit{I}(t)$ series, leading to a hierarchy of coherent structure intensities \citep{OsmanEA11,OsmanEA12}. The PVI statistic is constructed such that $\langle \mathit{I}^{2} \rangle = 1$, $\langle \mathit{I}^{4} \rangle$ is related to the kurtosis of the magnetic field increments, and powers of $\langle \mathit{I} \rangle$ scale in a manner connected with familiar diagnostics of intermittency.

\textit{Results}.---The scalar proton temperature $T_{p}$ and PVI statistic were each divided into $50 \times 50$ grids of logarithmically spaced bins in the ($\beta_{\parallel}$, $R$) plane, and the median values in each bin containing at least 50 measurements were computed. Figure 1a is (by construction) similar to Fig. 2 in \citep{MarucaEA11}, and solar wind observations are confined in the ($\beta_{\parallel}$, $R$) plane in a manner consistent with previous studies \citep[e.g.][]{HellingerEA06,BaleEA09,MatteiniEA07}. The highest $T_{p}$ values occur in regions typically associated with linear instability thresholds, and even at high $\beta_{\parallel}$ these areas are around 3--4 times hotter than the nearby $R=1$ plasma. However, the instabilities themselves are not thought to significantly heat the plasma \citep{Southwood93}, and so the heating mechanism has not been conclusively identified. 

Insight into the heating mechanisms contributing to these observations can be obtained by considering the PVI statistic. Figure 1b shows the median PVI values in the ($\beta_{\parallel}$, $R$) plane. The elevated $\mathit{I}$ values, which correspond to an increased likelihood of finding coherent magnetic structures such as current sheets within the observations, occur in almost exactly the same regions associated with enhancements in $T_{p}$. This result is completely consistent with earlier findings \citep{OsmanEA11} that samples of stronger PVI events systematically produce conditional probability distributions with higher mean proton temperature. Indeed, the same procedure also produces conditional distributions of electron temperature and electron heat flux, all with increased average values. However, it was not previously known that the most intense PVI events are preferentially found in plasma with high values of temperature anisotropy. This suggests coherent structures dynamically generated by MHD turbulence could be responsible for the anisotropic solar wind heating shown in Fig. 1a and observed by \citep{LiuEA06,MarucaEA11}. 

The relationship between PVI and $T_{p}$ is further investigated by dividing both into a $25 \times 25$ grid of logarithmically spaced bins, and computing the joint probability distribution $p(\mathit{I},T_{p})$ by dividing the number of observations in each bin $n$ by the total number of observations $N$. Figure 2a shows the proton temperature as a function of PVI, and the white dots are the most likely value of $T_{p}$ in each PVI bin. There is a clear dependence of proton temperature on PVI, and the most significant enhancements in $T_{p}$ are associated with PVI events above a threshold of around $\mathit{I} > 3$. These events correspond to non-Gaussian coherent current sheet structures \citep{GrecoEA09}, where the most intense are candidate magnetic reconnection sites \citep{ServidioEA11}. 

\begin{figure}[h]
\includegraphics[width=8.5cm]{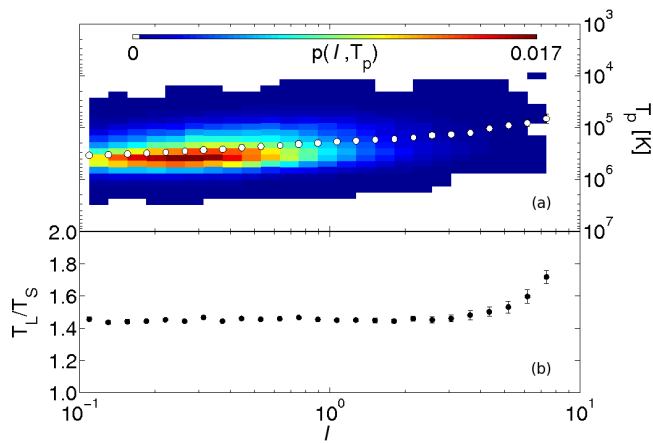}
\caption{Joint probability distribution between (a) proton temperature and PVI, where the most likely values of $T_{p}$ in each PVI bin are represented by white dots. There is a clear link between enhanced $T_{p}$ and strong PVI events. The most likely (b) temperature anisotropy value for each PVI bin, where $T_{L}$ ($T_{S}$) is the largest (smallest) of $T_{\perp}$ and $T_{\parallel}$. The most intense PVI events are associated with the greatest departures from temperature isotropy.} 
\label{Fig:anisoPDF}
\end{figure}

A link between the most intense PVI events and enhancements in $T_{p}$ has been established \citep{OsmanEA11,OsmanEA12}. However, it is unknown if a similar link exists between PVI and temperature anisotropy in the solar wind. Here we define $T_{L}$ and $T_{S}$ such that $T_{L}/T_{S} = R$ when $R \geq 1$ and $T_{L}/T_{S} = R^{-1}$ when $R < 1$. Figure 2b plots the most likely values of $T_{L}/T_{S}$, and the associated standard errors, for each of the 25 logarithmically spaced PVI bins. The temperature anisotropy remains almost constant until around $\mathit{I}=3$ when it starts to increase with stronger PVI events. This is consistent with temperature anisotropy being inhomogeneous and concentrated in the vicinity of coherent structures. These results are corroborated by Vlasov-Maxwell simulations, where temperature anisotropy is found to be enhanced in regions of strong magnetic stress between magnetic vortices \citep{ServidioEA12}. 

Temperature anisotropy near coherent structures could manifest through the preferential heating or cooling of protons either parallel or perpendicular to the mean magnetic field. Figure 3 shows PDFs of $T_{\parallel}$ and $T_{\perp}$ conditioned on PVI value, where $\mathit{I} \geq 0$ is the entire dataset, $\mathit{I} \geq 1$ removes low value fluctuations, $\mathit{I} \geq 3$ only retains non-Gaussian structures, and $\mathit{I} \geq 5$ contains only the most highly inhomogeneous structures including current sheets. Plasma with the greatest proportion of intense coherent structures has the highest most probable parallel and perpendicular temperatures. However, this heating appears in anisotropic proportions in the vicinity of the most non-Gaussian structures. Therefore, this suggest the presence of dynamical heating mechanisms within or nearby these coherent structures that can heat and accelerate protons both parallel and perpendicular to the magnetic field direction.

\textit{Disscussion}.---The results detailed here suggest distinctive kinetic signatures observed in solar wind plasma are associated statistically with coherent magnetic structures such as current sheets, which are connected to the intermittency properties of MHD turbulence. These kinetic effects include plasma heating and temperature anisotropy. The elevated proton temperature is due to broadening of the underlying VDF, which could be either a consequence of energy dissipation or particle energization, or a combination of both. Furthermore, plasma at the extremes of the accessible ($\beta_{\parallel}$, $R$) parameter space which is bounded by linear instability thresholds is not only likely to be hotter, but also likely to contain stronger magnetic discontinuities (as measured by higher PVI). Indeed, the pattern of median PVI values in this plane closely resembles the contours of median temperature.

\begin{figure*}[t]
\includegraphics[width=17cm]{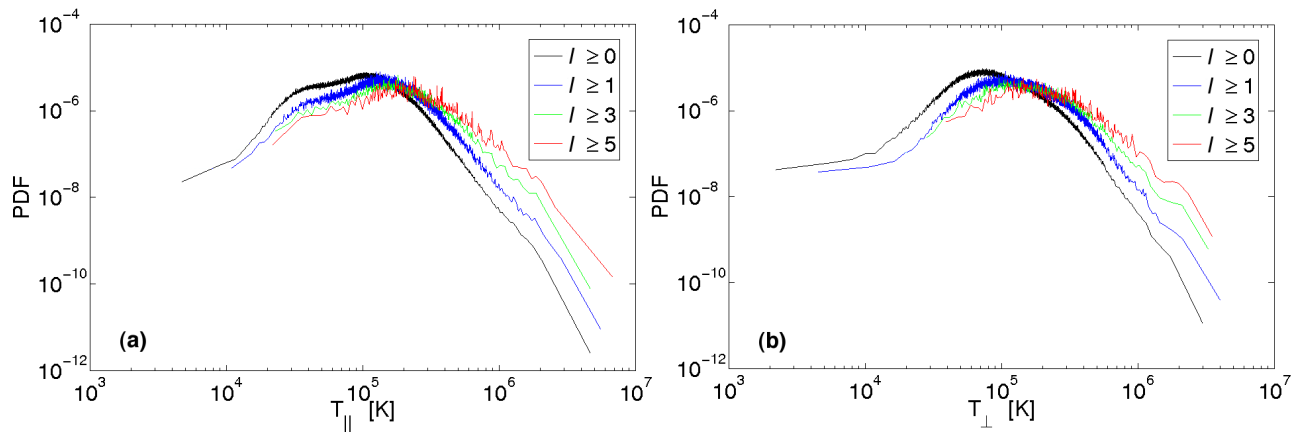}
\caption{PDFs of the (a) parallel and (b) perpendicular proton temperatures, where each corresponds to a different range of PVI values. In both cases, the strongest PVI events are associated with elevations in temperature.} 
\label{Fig:ppPDF}
\end{figure*}

We are led to the conclusion that kinetic effects in the solar wind plasma are inhomogeneous, being concentrated at the most intense coherent structures which are related to the intermittent properties of turbulence. However, the exact nature of this relationship between plasma turbulence and kinetic effects is unclear. It is known that MHD turbulence dynamically generates coherent structures, and these are sites of enhanced heating as in Fig. 2a \citep{OsmanEA11,OsmanEA12} and elevated temperature anisotropy as in Fig. 2b \citep{ServidioEA12}. This turbulence produced temperature anisotropy could then move the plasma to marginal instability, driving the growth of the firehose and mirror instabilities at the appropriate boundaries. Alternatively, the growth of these instabilities might result in the generation of coherent structures which heat the surrounding plasma anisotropically, thus driving it back to marginal stability. Indeed, there could be a complex feedback mechanism underlying the interaction between turbulence and these kinetic phenomena. However, the limiting regions of the ($\beta_{\parallel}$, $R$) plane seem to be roughly as well correlated to enhancements in the PVI statistic as they are to linear Vlasov instability thresholds. Therefore, consideration must be given to the possibility that an explanation to our observations need not include effects deriving from linear Vlasov theory.

There is evidence supporting the link between turbulence generated coherent structures and kinetic phenomena such as temperature anisotropy and plasma heating. This motivates further study to understand the detailed plasma physics underlying these connections. In particular, examination of dynamical activity associated with current sheets such as magnetic reconnection. For example, test particle scattering in MHD simulations shows most energetic protons are accelerated in or near current channels due to interaction with the nearby inhomogeneous electric field. These particles are not only accelerated to suprathermal speeds, but rapidly form anisotropic velocity distributions with speeds perpendicular to the mean magnetic field enhanced relative to speeds parallel \citep{DmitrukEA04}. There have also been indications of anisotropic temperatures near current sheets in 2.5D kinetic hybrid \citep{ParasharEA09} and hybrid Vlasov \citep{ServidioEA12} plasma turbulence simulations. Hence, there is strong impetus to understand dissipative and heating processes that operate non-uniformly in space. This should supplement the standard approach that focuses on linear models of kinetic dynamics within a uniform homogeneous plasma.

This work was supported by STFC, NSF Solar Terrestrial Program under grant AGS-1063439, NSF SHINE Program AGS-1156094, and by NASA under the Heliophysics Theory Program grant NNX11AJ44G and the Guest Investigator Program grant NNX09AG31G. The authors acknowledge useful conversations with J.C. Kasper, B.A. Maruca, L. Mattini, R.T. Wicks, T.S. Horbury, and K. Kiyani. K.T.O is a member of ISSI team 185.

\end{document}